\documentclass{aastex6}

\usepackage[utf8]{inputenc}

\shorttitle{DF~Cyg}
\shortauthors{Vega et al. 2017}

\newcommand{\longper}{$795 \pm 5$}
\newcommand{\shortper}{$49.84 \pm 0.02$}

\begin{document}

\submitted{Accepted for publication in The Astrophysical Journal}
\received{2017 January 13}
\accepted{2017 March 13}

\title{Evidence for Binarity and Possible Disk Obscuration in {\it Kepler} Observations of the Pulsating RV Tau Variable DF Cygni}

\author{Laura D. Vega\altaffilmark{1,2}, Keivan G. Stassun\altaffilmark{1,2}, Rodolfo Montez Jr.\altaffilmark{3}, Patricia T. Boyd\altaffilmark{4},
Garrett Somers\altaffilmark{1,5}}

\altaffiltext{1}{Department of Astrophysics, Vanderbilt University, Nashville, TN, USA}
\altaffiltext{2}{Department of Physics, Fisk University, Nashville, TN, USA}
\altaffiltext{3}{Smithsonian Astrophysical Observatory, Cambridge, MA, USA}
\altaffiltext{4}{NASA Goddard Space Flight Center, Greenbelt, MD, USA}
\altaffiltext{5}{VIDA Postdoctoral Fellow}

\begin{abstract}
The {\it Kepler} light curve of DF~Cyg is unparalleled in precision and cadence for any RV Tau star to date spanning a baseline of $\sim$4 years and clearly displaying the signature pulsating behavior of alternating deep and shallow minima as well as the long-term trend indicative of an RVb-type variable. We measure DF Cyg's formal period (the time interval between two successive deep minima) to be \shortper~days. The trend in the arrival times emulates that of the long-term period. There appear to be precisely 16 deep+shallow minima cycles in one long-term cycle, suggesting a long-term cycle period of $\approx$\longper~days. We argue that binarity may naturally explain the long-term periodicity in DF Cyg. The spectral energy distribution of DF Cyg features an infrared excess indicative of a disk possibly linked to a binary companion. From a recent {\it Gaia} parallax measurement for DF Cyg, we calculated that it has a distance of $990 \pm 372$~pc and a physical radius of $R_\star = 10.3 \pm 3.8$ R$_\odot$. From kinematics and geometric arguments, we argue that the most likely interpretation for the decrease in flux from the long-period maximum to the long-period minimum, as well as the reduction of the short-term pulsation amplitude, is caused by an occulting body such as a disk surrounding DF Cyg and its binary companion. 
\end{abstract}

\textbf{}

%----------------------------------------------------------------------------------------
% 	Section:  INTRODUCTION
%----------------------------------------------------------------------------------------
\section{Introduction}
\label{sec:intro}
RV Tau variable stars, named after their prototype RV Tauri, are luminous large-amplitude pulsating supergiants \citep[General Catalog of Variable Stars (GCVS);][]{Samus2009} located at the brightest part of the population Type II Cepheid instability strip in the Hertzsprung--Russell diagram \citep{Wallerstein2002}. The colors of RV Tau stars are in phase with their variation in brightness, displaying spectral types between F and G at maximum light and K to M at minimum light \citep{Samus2009}. The large luminosities and large IR excess due to circumstellar dust exhibited by RV Tau variables led to their classification as post-asymptotic giant branch (post-AGB) objects, suggesting they can be planetary nebula progenitors \citep{Jura1986}. 

The main characteristic of RV Tau variables is a double wave of alternating deep and shallow minima in their light curve. The combination of deep and shallow minima has a ``formal'' period (the time interval between two successive deep minima) that ranges between 30 and 150 days \citep{Samus2009}. This short-term feature of repeating deep and shallow minima has been argued \citep{Gerasimovic1929} to represent two pulsation modes simultaneously being excited in a 2:1 resonance  
\citep[for a detailed description, see][]{Pollard1996}. In addition to this short-term behavior, RV Tau variables are divided into two photometric subclasses based on longer-term variability. RVa stars maintain a relatively constant mean magnitude throughout their alternating deep and shallow minima. RVb stars, on the other hand, show an additional long-period variation on a timescale of 600--1500 days in mean magnitude with amplitudes that can reach up to 2 mag in $V$ \citep{Samus2009}.

A major outstanding question in the study of RV Tau stars is the origin of these two classes based on the presence of long-period variation (in their light curves) or lack thereof. Various studies \citep[e.g.,][]{Vanwinckel1999,Maas2002,Deruyter2005RVTau,Gezer2015,Manick2016} have argued that RV Tau stars generally can be understood if they possess binary companions. \citet{Vanwinckel1999} compared various observed characteristics of RV Tau stars with post-AGBs in known binary systems and found that a high fraction of RV Tau stars were similar to their binary post-AGB counterparts. They further suggested that the viewing angle of a circumstellar dust torus determined the long-term variability class, with RVa stars being seen pole-on and the RVb stars seen edge-on. The RVb long-term variation then is due to periodic extinction by orbiting circumstellar dust. Based on IR excess in the spectral energy distributions (SEDs). indicative of the dust disks of six RV Tau systems, \citet{Deruyter2005RVTau} found that the most likely explanation for the inferred structure and size of the disks is binarity. These studies further support the view that binary companions, and possibly also circumstellar disks, are important for understanding the RV Tau phenomena in general. A binary origin for the RV Tau phenomena would furthermore help to connect RV Tau stars as evolutionary precursors of planetary nebulae, as studies suggest that the asymmetric morphology in the majority of planetary nebulae is also due to binarity \citep[see, e.g.,][]{Demarco2009}. 

However, the question of what drives the RVa versus RVb long-term photometric behavior remains. \citet{Gezer2015} used {\it Wide-field Infrared Survey Explorer (WISE)} data to study all galactic RV Tau stars in the GCVS \citep{Samus2009} catalog by comparing them to post-AGB objects. They found that all RVb stars in their sample exhibited IR excesses and found them only among the disk sources, whereas some but not all of the RVa stars exhibited such IR excesses. More importantly, they found that all confirmed binaries in their sample were disk sources. Thus, there is evidence that the presence of a dust disk is most strongly connected to the presence of RVb long-term photometric variations in RV Tau stars. 

If dust disks cause the signature RVb long-term variation through orbital modulation of the obscuration along the line of sight, then such orbital modulation might be the result of a binary companion, in which case one might expect the modulation to be related to the binary orbital period. 
In fact, there is evidence for this scenario in RVb stars such as U Mon, an example of a confirmed binary with a dust disk \citep{Gezer2015}, that has a long-term photometric period of 2475 days \citep{Percy1991} and agrees with the binary orbital period of 2597 days found by \citet{Pollard1995}. 

Since the discovery of DF Cyg \citep{Harwood1927}, this archetypal RVb star, which varies in magnitude between $\approx$13--10.5 mag \citep{Bodi2016}, has remained a mystery. Although there have been some early indications of possible binarity, the radial velocity signature of a binary companion has been inconclusive. For example, \citet{Joy1952} reported a range in the radial velocities of $45 {\rm ~km~s}^{-1}$, measured in 10 plates obtained at Mt.~Wilson Observatory, but with a large uncertainty. Moreover, \citet{Gezer2015} classified the SED of DF~Cyg as ``uncertain" with regards to the evidence for a dust disk. 

Fortunately, however, DF~Cyg was included in the recent {\it Kepler} mission, providing an unprecedented high-quality light curve spanning $\approx$4 years (see Figure~\ref{fig:fulllc}). {\it Kepler} began observing DF~Cyg at the start of a shallow minimum followed by a total of 29 full double-wave (deep + shallow minima) cycles, covering almost two full cycles of an $\sim$800 day long-term RVb period. In addition, \citet{Gorlova2013} recently reported long-term radial velocity variations with a period of $\sim$775 days for DF~Cyg. Finally, DF~Cyg was included in the first data release of parallax observations by {\it Gaia} \citep[][]{Gaia2016a,Gaia2016b}. Together, the {\it Kepler} light curve and these recent radial velocity and parallax measurements permit a new comprehensive re-evaluation of this important RV Tau system. 

In this paper, we present updated measurements of the periodicities in the DF~Cyg light curve and show conclusively that DF~Cyg possesses a large IR excess indicative of a dust disk. In addition, we present timescale arguments to suggest that the long-term RVb variations in DF~Cyg are consistent with a disk obscuration hypothesis on the binary orbital timescale. Section~\ref{sec:data} summarizes the {\it Kepler} light curve data that we use in our analysis and Section~\ref{sec:analysis} describes our analysis procedures and the main results that emerge from them. We discuss our interpretation of the results in Section~\ref{sec:disc}, and conclude with a summary of our findings in Section~\ref{sec:summary}.

%----------------------------------------------------------------------------------------
% 	Section:  DATA
%----------------------------------------------------------------------------------------
\section{Data: {\it Kepler} Time Series Observations}
\label{sec:data}

DF~Cyg's brightness variability and that for more than $\sim$150,000~target stars were simultaneously monitored by {\it Kepler} through a long time baseline spanning $\sim$4 years. {\it Kepler} was designed to survey the Cygnus--Lyra star field region with a 105 degree\textsuperscript{2} field of view. A total of 18 observation quarters each lasting $\sim$90 days were obtained from 2009 May 2 to 2013 April 9 (Q0, Q1, Q8, and Q17 were approximately 10 days, 33 days, 67 days, and 32 days, respectively). In 2013 May, {\it Kepler} lost the second of its four reaction wheels, which ended the continuous monitoring mission in the field. However, the observations obtained by {\it Kepler} provide space-based light curves with the highest photometric accuracy and uninterrupted coverage compared to any ground-based observatory.

{\it Kepler} had two observing modes: the short-cadence mode and the long-cadence mode, which took an image (every $\sim$1 minute or every $\sim$30 minutes, respectively) for the duration of the entire mission \citep{Hartig2014}. There are 0.9 -- 3 day gaps in the data from the transition of one observing quarter to the next because {\it Kepler} was required to rotate every three months to maintain direct sunlight on the solar arrays, optimizing their efficiency. Most of the targets fell on a different CCD channel at every observation quarter, and their point-spread functions were distributed among different neighboring pixel apertures. As a result, flux discontinuities exist between quarter-to-quarter transitions. 

We retrieved the {\it Kepler} ``simple aperture photometry" (SAP), long-cadence mode data of DF Cyg from the Mikulski Archive for Space Telescopes (MAST). The data are highly sampled in time with a total of $\sim$69,778 data points throughout the 18 observation quarters. Data were collected on day $\sim$120 of the mission and ended at day $\sim$1591, summing to a total of $\sim$4 years of observations.

We use the SAP data as is; however, there were noticeable systematic discontinuities in flux throughout the light curve, due to quarter-to-quarter transitions. We shifted quarters 0--3, 9, and 12 to reduce these discontinuities to best represent the signal of the DF~Cyg. These changes were relatively small but helped reduced the error on our measured long-term period value. For an in-depth description on the removal of instrumental effects on {\it Kepler} data; see, e.g., \cite{Hartig2014}.

%----------------------------------------------------------------------------------------
% 	Section:  ANALYSIS
%----------------------------------------------------------------------------------------
\section{Analysis and Results}
\label{sec:analysis}

In this section we present the results of our analysis of the {\it Kepler} light curve observations of DF~Cyg. We give a general characterization of the overall light curve features. We obtain precise measurements of both the short-term period and long-term period. Finally, we present the results of our analysis of the available broadband flux measurements of DF~Cyg, which we use to construct an SED and to estimate the radius of DF~Cyg as well as to assess the evidence for a dust disk in the system.

\subsection{General Features of the DF~Cyg Light Curve}
\label{sec:lightcurve}

\begin{figure}[!ht]
\centering
\includegraphics[width=\textwidth,height=\textheight,keepaspectratio]{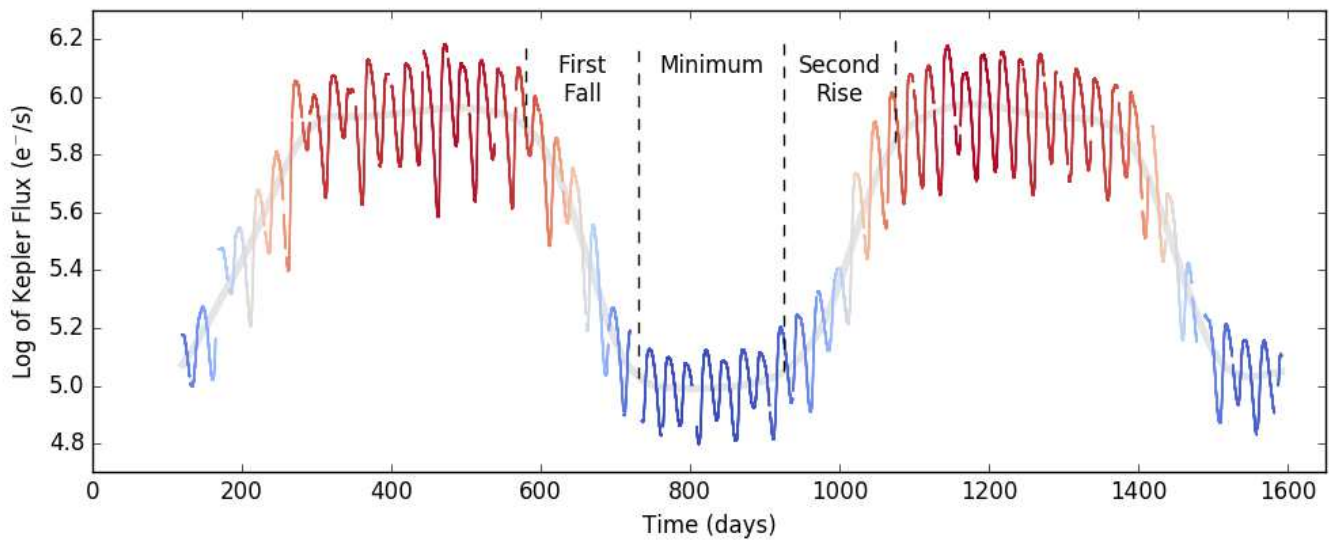}
\caption{{\it Kepler} light curve for DF Cyg with a spline model (in gray) underlying the long-period trend. The data points are color-coded according to the value of the spline model, where red means maximum light and blue means minimum light. The dashed vertical lines indicate the approximate transitions of light to and from minimum light. Note that the short-term oscillation amplitudes decrease at long-period minimum.}
\label{fig:fulllc}
\end{figure}

Figure~\ref{fig:fulllc} displays the {\it Kepler} light curve of DF~Cyg. 
The variations present in the light curve show the characteristic RV Tau alternating minima and a long-period variation in the mean brightness, which are indicative of the RVb classification. {\it Kepler} observed a complete cycle of the long-period variation, starting as DF~Cyg emerged from a long-period minimum and ending as DF~Cyg entered a second long-period minimum. As a result, the {\it Kepler} light curve features two maximum states that bracket one complete minimum state.

\begin{figure}[!ht]
\centering
\includegraphics[width=\textwidth,height=\textheight,keepaspectratio]{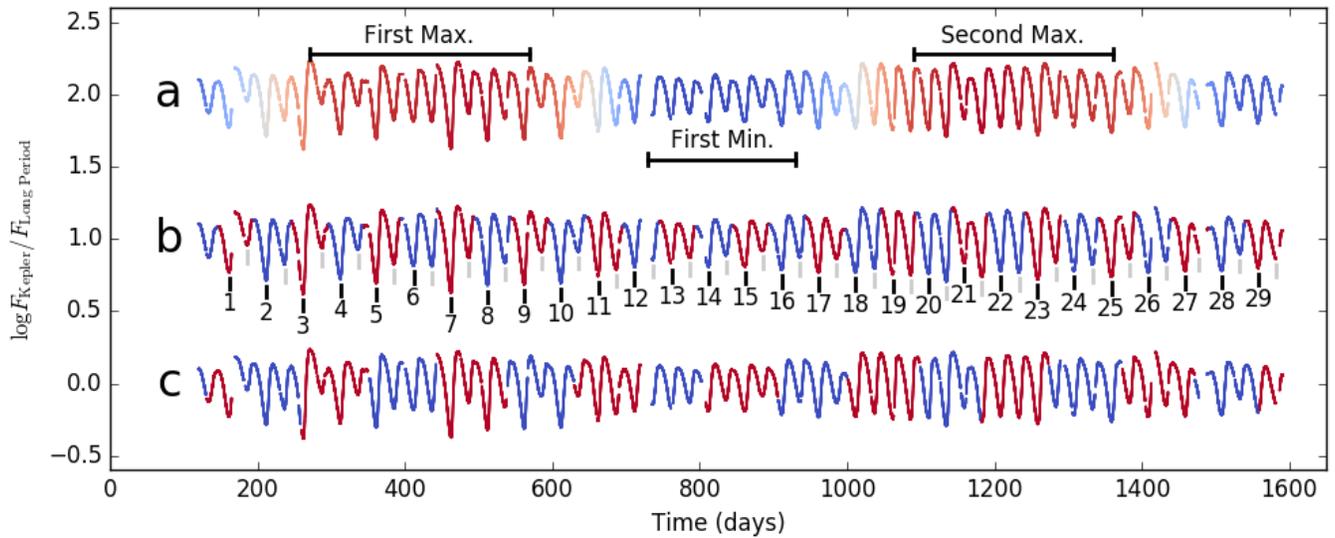}
\caption{{\it Kepler} light curve for DF~Cyg with long-term behavior removed. From top to bottom, we  color-code the data according to (a) the value of the long-term behavior spline model, (b) alternating colors based on our double-wave cycle breakdown (there are a total of 29 deep + shallow minima cycles), and (c) alternating colors based on {\it Kepler} quarters. Location of the extrema of the long (a) and short (b) behaviors are also depicted.}
\label{fig:cycleslc}
\end{figure}

The {\it Kepler} light curve includes 29 full deep + shallow minima cycles, which we have labeled in Figure~\ref{fig:cycleslc}. Both of the long-term maxima are flat with the mean brightness roughly lasting $\sim$300$\pm$70 days each. The long-term minimum is $\sim$200$\pm$70 days in duration. The long-term brightness rise and fall times are $\sim$150$\pm$70 days each. These general timescales are summarized in Table~\ref{T-observation}. The overall flux decreases by $\sim$90\% from the long-period maximum to the long-period minimum. During the long-period maxima, we find that the average peak-to-peak amplitude of the short-period behavior is $\approx 9.2 \times 10^5$ flux units, while in the long-period minimum the average peak-to-peak amplitude is $\approx 7.9 \times 10^4$ flux units. Just as the overall flux decreases by 90\%,
the short-period oscillation amplitudes decrease by $\sim$90\% during the long-period minimum relative to the long-period maximum. In other words, the short-period oscillations are, fractionally, the same during both the long-period maxima and the long-period minimum states. \citet{Percy1993} similarly found that in U~Mon, RV~Tau, and DF~Cyg, the amplitudes of the short-period oscillations get smaller during the long-period minimum. Here, with DF~Cyg's {\it Kepler} light curve, we can clearly see the highly resolved short-period oscillations are $\sim$90\% much lower in amplitude during the long-period minima. In Section~\ref{sec:disc}, we return to these phenomena in the context of an occultation scenario for DF~Cyg.

\begin{table}[!ht]
\caption{Long-period Transitional Behavior}
\centering
\label{T-observation}
\begin{tabular}{lcccc}
\toprule
Event  & $T_{\rm start}$ & $T_{\rm end}$ (Days)	& Duration (Days) \\
\hline
First Rise\tablenotemark{a} & 125$\pm50$ & 275$\pm50$ 	    & 150$\pm70$\\
First Maximum & 275$\pm50$  & 575$\pm50$  	& 300$\pm70$\\ 
First Fall & 575$\pm50$  & 725$\pm50$  	& 150$\pm70$\\
First Minimum     & 725$\pm50$  & 925$\pm50$      & 200$\pm70$\\ 
Second Rise    & 925$\pm50$  & 1075$\pm50$     & 150$\pm70$\\
Second Maximum	& 1075$\pm50$ & 1375$\pm50$ 	& 300$\pm70$\\ 
Second Fall & 1375$\pm50$ & 1525$\pm50$ 	& 150$\pm70$\\
\hline
\end{tabular}
%}
\tablecomments{Approximate day on which transitions occur.}
\tablenotetext{a}{The first rise may be incomplete due to the time at which {\it Kepler} started observing.}
\end{table}

\subsection{Arrival Time Variations in the Deep and Shallow Minima}
\label{sec:minima}
\begin{figure}[!ht]
\centering
\includegraphics[width=4in]{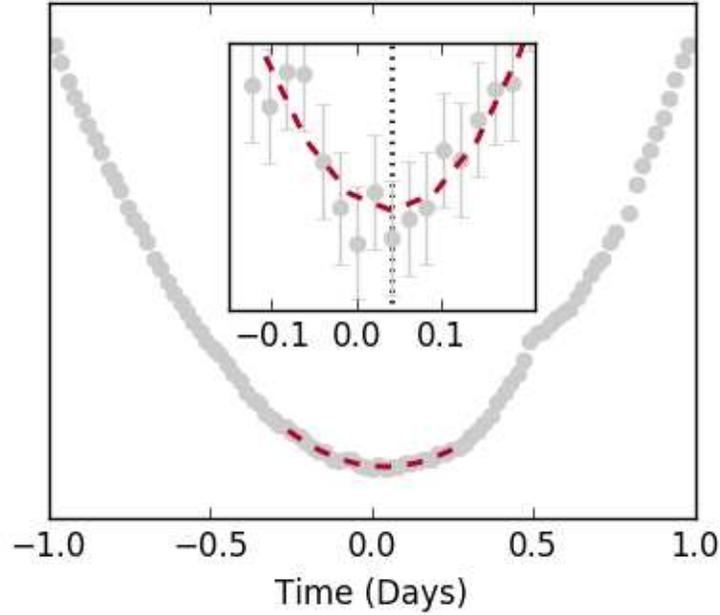}
\caption{Example of arrival time estimate on the deep minimum Cycle 17. The time of the light curve segment is normalized to the time of the data point with the minimum flux. The actual arrival time of the minimum as determined from the Gaussian fit (dashed red line) is indicated by the vertical dotted line shown in the inset zoom.
}\label{fig:gfits}
\end{figure}

\begin{figure}[!ht]
\centering
\begin{tabular}{cc}
\includegraphics[width=\textwidth,height=\textheight,keepaspectratio]{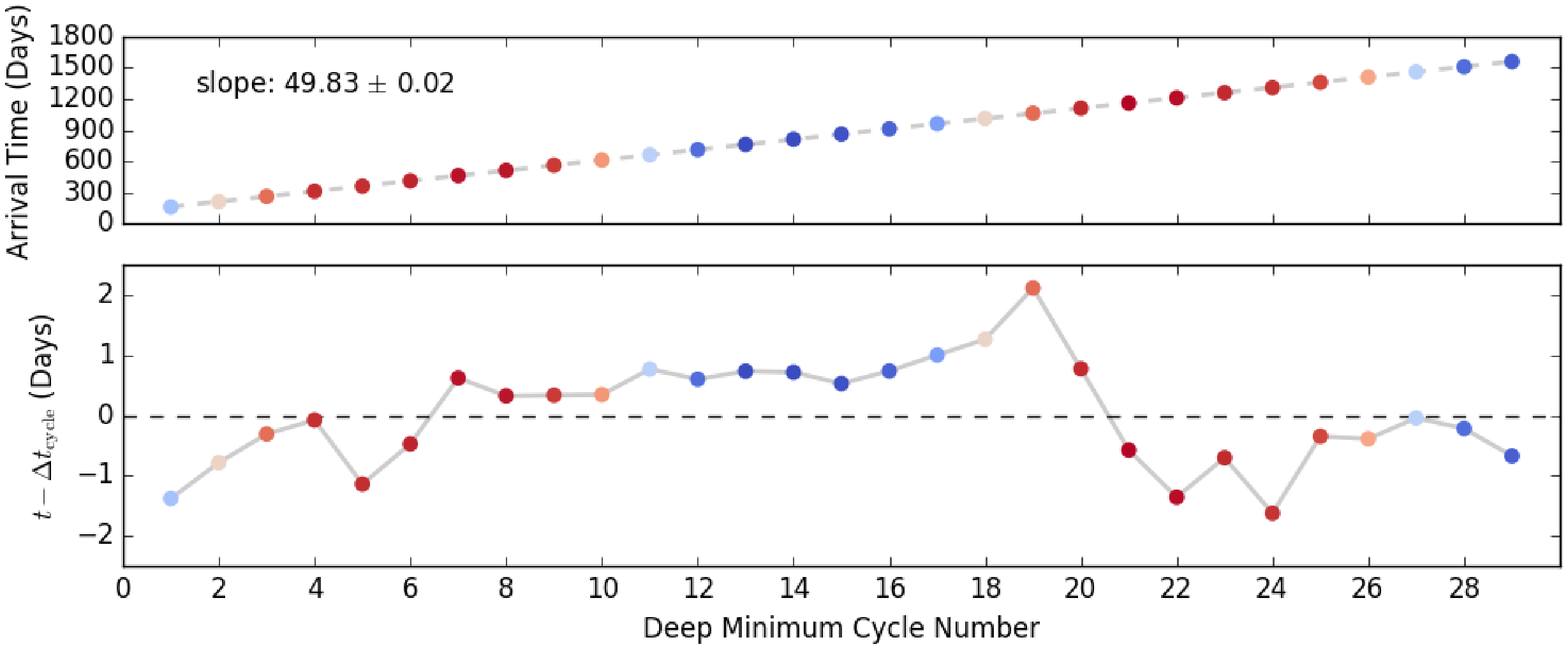}
\end{tabular}
\caption{Arrival time results. Top panel: the Gaussian-fit arrival times are plotted against the cycle number. The best-fit line to the data is indicated by the dashed line. Lower panel: the residuals of the arrival times and best-fit model are plotted against the cycle number, in the sense of $O-C$ (observed minus calculated). In both panels, the data points are color-coded according to the spline model of the long-period variation, as in Figure~\ref{fig:fulllc}. Errors on the arrival times and on the residuals are smaller than the data points.
}\label{fig:minfits}
\end{figure}

To verify the periodicity of the short-term behavior in DF~Cyg, 
and to assess any possible variability in the short-term periodicity, 
we sought to measure the ``arrival time" of each light curve minimum. 
Figure~\ref{fig:gfits} shows an example of a deep minimum, Cycle 17, observed by {\it Kepler}. We fit all 29 deep minima with a Gaussian function, where we used the centroid of the Gaussian, $\mu_{\rm min}$, and its associated error, $\mu_{\rm err}$, to characterize the arrival time of each minimum. Our best-fit arrival time values and associated errors for each deep minimum are presented in Table~\ref{deeparrivaltimes} and in Figure~\ref{fig:minfits} (top panel), where we show the measured arrival times of each deep minimum as a function of cycle number. We note that the error bars, $\mu_{\rm err}$, are smaller than the data points.

\begin{table}[!ht]
\caption{Deep Minima Arrival Time}
\label{deeparrivaltimes}
\centering
\scalebox{.9}{
\begin{tabular}{cccc}
\toprule
Cycle Number &Arrival Time (Days)	&Lag Time (Days)\\
\hline

1 & 161.628$\pm$0.003 & -1.43$\pm$0.020 \\
2 & 212.059$\pm$0.002 & -0.83$\pm$0.020 \\
3 & 262.372$\pm$0.001 & -0.35$\pm$0.020 \\
4 & 312.438$\pm$0.001 & -0.12$\pm$0.020 \\
5 & 361.214$\pm$0.001 & -1.18$\pm$0.020 \\
6 & 411.716$\pm$0.001 & -0.51$\pm$0.020 \\
7 & 462.645$\pm$0.002 & 0.58$\pm$0.020 \\
8 & 512.185$\pm$0.001 & 0.29$\pm$0.020 \\
9 & 562.033$\pm$0.001 & 0.30$\pm$0.020 \\
10 & 611.880$\pm$0.002 & 0.31$\pm$0.020 \\
11 & 662.138$\pm$0.001 & 0.73$\pm$0.020 \\
12 & 711.808$\pm$0.002 & 0.57$\pm$0.020 \\
13 & 761.780$\pm$0.006 & 0.71$\pm$0.021 \\
14 & 811.595$\pm$0.006 & 0.69$\pm$0.021 \\
15 & 861.244$\pm$0.005 & 0.50$\pm$0.020 \\
16 & 911.290$\pm$0.008 & 0.71$\pm$0.021 \\
17 & 961.392$\pm$0.007 & 0.98$\pm$0.021 \\
18 & 1011.488$\pm$0.003 & 1.24$\pm$0.020 \\
19 & 1062.175$\pm$0.002 & 2.09$\pm$0.020 \\
20 & 1110.671$\pm$0.001 & 0.75$\pm$0.020 \\
21 & 1159.160$\pm$0.003 & -0.59$\pm$0.020 \\
22 & 1208.217$\pm$0.001 & -1.37$\pm$0.020 \\
23 & 1258.706$\pm$0.001 & -0.72$\pm$0.020 \\
24 & 1307.624$\pm$0.002 & -1.63$\pm$0.020 \\
25 & 1358.736$\pm$0.002 & -0.36$\pm$0.020 \\
26 & 1408.529$\pm$0.002 & -0.40$\pm$0.020 \\
27 & 1458.711$\pm$0.002 & -0.05$\pm$0.020 \\
28 & 1508.374$\pm$0.006 & -0.22$\pm$0.021 \\
29 & 1557.755$\pm$0.005 & -0.68$\pm$0.020 \\

\hline
\end{tabular}
}
\tablecomments{Gaussian-fit arrival time for each deep minimum cycle and residuals plotted in Figure~\ref{fig:gfits}.}
\end{table}

The linear trend of the data indicates the overall consistency of the arrival times of the minima, with a best-fit slope of \shortper~days, which we take to be the short-term period for DF~Cyg. This is consistent with the period of 49.4~days (no quoted uncertainty) originally reported by \citet{Harwood1927}, as well as with the recent period reported by \citet{Bodi2016} of 49.9 days.

Subtracting the linear regression from the arrival times gives the residuals, which show systematic deviations, of the arrival times about the best-fit linear trend (Figure~\ref{fig:minfits}, bottom panel). The deviations are typically on the order of $\sim$0.5~days, with an amplitude of $\sim$2~days. These deviations are highly significant considering the typical uncertainty on the individual arrival times of 0.02~d. The pattern of the residuals is not strictly sinusoidal, but does undergo a sign flip (from negative to positive residual) around cycles 4--6 and again (from positive to negative residual) around cycle 21. 
The timescale between these sign flips, approximately 15--17 cycles, corresponds to a period in the range 747--847~days, given the average short-term cycle period of \shortper~days found above.

\subsection{Redetermination of the Long-term Periodicity} 
\label{sec:longterm}

\begin{figure}[!ht]
\centering
\includegraphics{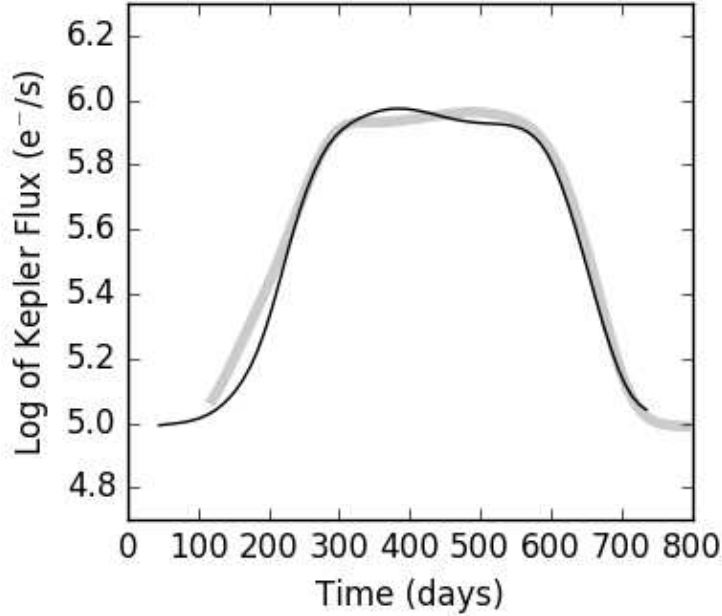}
\caption{Phasing the Long-period spline. The gray curve represents the first half of the long-period cycle determined from the spline model described in the text. The black curve represents the second half of the long-period cycle phased by 795-d. Phases greater than $\pm5$ days can be disregarded, giving the best-fit long-term period of \longper~days.}
\label{fig:phasingspline}
\end{figure}

DF~Cyg's {\it Kepler} light curve almost spans two full long-period cycles.
As an initial estimate, we divided the light curve into two parts at day 800 (Figure~\ref{fig:phasingspline}). We heavily smoothed both portions using a spline with a smoothing length of 1150 data points, and cross-correlated the two spline smoothing functions to get a best-fit long-period timescale of \longper~days. This period is consistent with previous a estimate \citep{Harwood1937}, and is within 2$\sigma$ of the 779.6$\pm$0.2~day value reported by \citet{Bodi2016} from 50 years of AAVSO measurements.

We note that a period of $\sim$800~days is almost exactly 16 times the short period we found above of \shortper~days, which would correspond to a long period of 797.4$\pm$0.3~days. Long-to-short period ratios of this order are generally observed in other RVb stars \citep[][]{Tsesevich1975,Percy1993}. When we phased the {\it Kepler} light curve by this long-period value (Figure \ref{fig:phasingspline}), the arrival times of the short-period oscillations align well. However, the amplitudes do differ, indicating that the variations in the system are not strictly periodic in every detail. At the same time, the long-period behavior does show broad repeatability at this period, especially in the alignment of the rises to long-period maxima and the descents toward long-period minima. In addition, the flux at long-period maxima are both flat. The flat long-period maxima and their different durations resemble the RVb binary, U Mon \citep{Percy1993}. Overall, we remark that the long-term periodic behavior is reminiscent of an eclipsing binary. We return to the idea of a periodic occultation hypothesis in Section~\ref{sec:disc}.

Finally, we noticed that at both of the long-period maxima, the pattern of deep and shallow minima becomes temporarily interrupted. Specifically, the shallow minimum of Cycle 5 and the subsequent deep minimum of Cycle 6 (at day $\sim$400) have roughly the same absolute flux and are not as readily distinguishable as deep versus shallow. The same occurs for the shallow minimum of Cycle 21 and the subsequent deep minimum of Cycle 22 (at day $\sim$1200; Figure~\ref{fig:cycleslc}). These temporary interruptions coincide with the sign reversals in the arrival time residuals observed in Section~\ref{sec:minima} and Figure~\ref{fig:minfits}, a behavior separated by $\sim$800~days.

\subsection{SED: Stellar Radius and Dust Disk}
\label{sec:sed}
\begin{figure}[!ht]
\centering
\includegraphics[width=0.65\textwidth,height=0.55\textheight,keepaspectratio]{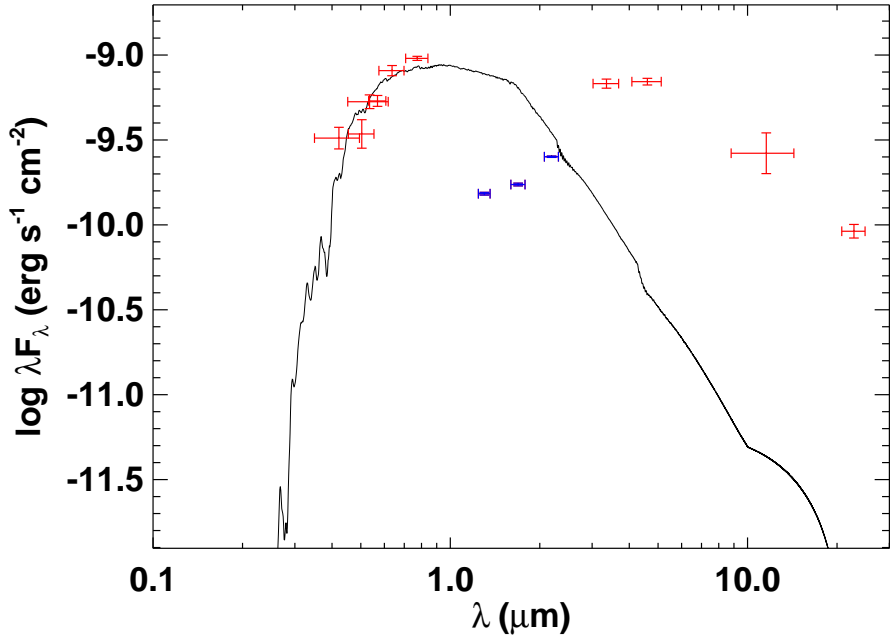}
\caption{SED with our best-fitting stellar atmosphere model to the optical photometry (black line). Measured fluxes are shown as red crosses where the vertical bars represent the measurement uncertainties and the horizontal bars represent the width of the passband. The aberrant 2MASS $JHK_S$ photometry (blue data points) is due to those observations having been taken during an RVb dimming of DF~Cyg. The {\it WISE} measurements reveal a clear and strong infrared excess, typically indicative of a dusty disk (see the text).}\label{fig:sed}
\end{figure}

We gathered the available broadband photometric data for DF Cyg from {\it GALEX}, {\it Tycho-2}, {\it APASS}, {\it 2MASS}, and {\it WISE} to produce the observed SED spanning the wavelength range 0.2--20~$\mu$m as shown in Figure~\ref{fig:sed}. We used an adopted effective temperature of $T_{\rm eff} = 4840$~K \citep{Giridhar2005,Brown2011,Bodi2016}. The 2MASS fluxes are clearly anomalous, and we confirmed from the timestamps in the 2MASS catalog that this is due to the 2MASS observations of DF~Cyg having been taken near minimum brightness of the star \citep{Skrutskie2006}. In addition, the {\it WISE} bands exhibit a clear excess. Therefore, we attempted a fit to only the portion of the SED at $\lambda < 1$~$\mu$m. The UV to optical portion of the SED can be explained with a Kurucz stellar atmosphere model (we adopt solar metallicity for simplicity) with a best-fit extinction of $A_V = 0.61 \pm 0.06$ mag, implying a distance of $\sim$1.3~kpc for this sightline based on the 3D dust maps of \citet{Green:2015}. Indeed, the newly available {\it Gaia} parallax of $\pi = 1.01 \pm 0.38$~mas puts DF Cyg at $990 \pm 372$~pc.

Integrating the model SED for the DF~Cyg photosphere (black curve in Figure~\ref{fig:sed}) gives a (dereddened) bolometric flux at Earth of $F_{\rm bol} = 1.71 \pm 0.05 \times10^{-9} {\rm ~erg~s}^{-1} {\rm ~cm}^{-2}$. With $T_{\rm eff} = 4840$~K (see above), this yields an angular radius of $\Theta = 0.048 \pm 0.001$~mas. Using this calculated angular radius with the {\it Gaia} distance in turn yields a physical radius for DF~Cyg of $R_\star = 10.3 \pm 3.8$ R$_\odot$, consistent with an evolved status for DF~Cyg. We note that this radius is somewhat smaller than is typically found for RV~Tau stars using period--luminosity--color relations \citep[e.g.,][]{Manick2016}.

In addition, there is a large IR excess evident from the {\it WISE} data, consistent with that expected from a dusty circumstellar disk (\citep{Gezer2015}; disk model discussed in Section~\ref{sec:disc}). 
The IR excess was not conclusive in the study of \citet{Gezer2015} because the {\it WISE} data alone exhibit a combination of [3.4]$-$[4.6] and [12]$-$[22] colors that placed DF~Cyg just at the edge of their ``disk" criterion. Here, the inclusion of the full SED makes the detection of a large IR excess indicative of a disk unambiguous.
In comparison, the 2MASS $JHK_S$ measurements appear anomalously low. However, we note that the 2MASS observations were obtained during a long-period minimum state (1998 June 13), with a phase corresponding to day $\sim$800 in Figure~\ref{fig:fulllc}. These measurements are further discussed in Section~\ref{sec:disc}.

%----------------------------------------------------------------------------------------
% 	Section:  DISCUSSION
%----------------------------------------------------------------------------------------
\section{Discussion: Disk Occultation Scenario for the Long-period Behavior}
\label{sec:disc}
In this section, we discuss the stellar parameters we obtained for DF~Cyg, such as periodicity, the residual trend from the short-period minima arrival times, as well as its radius and the confirmation of its IR excess. We develop a scenario in which DF~Cyg and its binary companion undergo periodic occultations by a dusty circumbinary disk.

\subsection{Evidence for a Binary Companion with an 800 Day Period}
There is now strong evidence to suggest the presence of a binary companion orbiting DF~Cyg with a period of $\sim$800~days. 
\citet{Gorlova2013} reported preliminary results from a long-term radial velocity monitoring program of $\sim$70 supergiants that included post-AGB stars with disk detections. They found DF~Cyg to be one of their best binary candidates, with a period of $\sim$775~days (no uncertainty reported). Some of the other candidates include EP~Lyr ($\sim$1100~days), R~Sge ($\approx$1159~days), and RV~Tau ($\approx$1210~days). All of these are RVb stars with previously recorded long-term photometric periods ranging from $\sim$1100--1200 days \citep[e.g.,][]{LloydEvans1985,Gielen2009}. 

In the case of DF~Cyg specifically, if we assume a circular Keplerian orbit for the binary companion and adopt a typical post-AGB mass of 0.6~$M_\odot$ for the mass of DF~Cyg \citep{Weidemann1990}, we obtain a range of semi-major axis values depending on the assumed mass of the companion. For example, for companion masses in the range 1--3~$M_\odot$, we obtain a range of semimajor axis $a \approx$ 1.5--2.5~AU. 

Importantly, all of the previously identified RVb stars with binary companions appear to also possess dusty disks, based on their SEDs \citep{Gorlova2013,Gezer2015}. In addition, \citet{Manick2016} reported that all six of the RV Tau stars in their radial velocity study have disks and have binary companions with orbital periods of 650--1700~days and eccentricities of 0.2--0.6. Therefore, in both of these larger samples of RV Tau stars, there appears to be a connection between the presence of a binary companion, the signatures of dusty disk material, and the periods of both the binary orbit and the long-period photometric ``RVb" variation. 

Similarly, our results for the behavior in the short-period oscillation arrival times (Section~\ref{sec:analysis}) suggest that DF~Cyg undergoes some sort of perturbation with a period comparable to the long-period ``RVb" oscillation of \longper~days, which is itself comparable to the binary companion period of $\sim$775~days reported by \citet{Gorlova2013}. A phenomenon related to heartbeat stars \citep{Thompson2012}, whose tidal distortions at periastron introduce periodic variations in the light curve, could be related to the temporary interruptions of the deep and shallow minima. 

This confluence of variability timescales---all around $\sim$800 days---motivates us to consider a unifying interpretation involving the binary companion and a dusty disk in the circumbinary environment of DF~Cyg, as we now discuss.

\subsection{Disk Occultation Scenario} \label{sec:disk}

We begin by ruling out the possibility of having only a simple eclipsing binary star scenario, in which the $\sim$90\% decrease in light during the long-period minimum is caused by DF~Cyg eclipsing a hot, luminous companion that contributes $\sim$90\% of the total system light. First, the duration of the long-period ``RVb" dimming is very long---about half of the orbital period. It is very difficult to construct an orbital configuration in which a companion star can spend half of the orbit traversing behind DF~Cyg. Perhaps this could occur if the orbit is highly eccentric; however, the orientation of the orbit would have to be fortuitously aligned with our line of sight, such that the companion is being blocked at apastron to account for the very long duration it would then spend in eclipse. 

Second, in this scenario, we would expect a secondary eclipse to occur at periastron as the companion passes in front of DF~Cyg. However, we do not see any signs of a secondary eclipse in the light curve near the phases of the long-period maxima. Finally, since the short-period oscillations are observed to decrease by $\sim$90\% in amplitude during the long-period minimum, it must almost certainly be the case that it is DF~Cyg that is principally occulted during long-period minima. This instead requires that $\sim$90\% of the stellar disc of DF~Cyg is obscured by a very large, opaque screen. 

If we instead interpret the long-period changes in average flux to be due to ingresses, eclipse, and egresses of an opaque body passing in front the DF~Cyg + companion system, we can attempt to characterize the occulting body. Given that DF Cyg exhibits a clear IR excess indicative of a dusty disk somewhere in the system, one interpretation is that the occulting screen is a feature in a disk around DF~Cyg itself or else a disk around the companion star. Therefore, we first consider the approach laid out in \citet{Rodriguez2013} for the occultation of a star by an orbiting disk. The extent of the screen can be obtained by calculating its transverse velocity, using our empirically obtained radius for DF~Cyg and the duration of ingress (or egress). We calculated the size, $w$, of the occulting screen from the screen's transverse velocity, $v_{\rm T}$, and the amount of time, $t_{\rm eclipse}$, given by the sum of the ingress duration and the total eclipse duration, 
$t_{\rm eclipse} = w/v_{\rm T}$. The transverse velocity is itself related to the size of DF~Cyg and the ingress duration, $t_{\rm ingress}$, by $v_{\rm T} = 2 R_\star / t_{\rm ingress}$, where $R_\star$ is the radius of DF~Cyg. Thus, $w = 2 R_\star t_{\rm eclipse}/t_{\rm ingress}$. With an observed $t_{\rm ingress}$ = 150 $\pm$ 70 days and observed $t_{\rm eclipse}$ = 350 $\pm$ 70 days, we obtain $v_{\rm T}$ = 1.11 $\pm$ $0.66 {\rm ~km~s}^{-1}$ and $w=$ 0.22 $\pm$ 0.14 AU.

We considered two possibilities for the location of the dusty disk occulting DF~Cyg and producing dimmings, in the context of the above scenario from \citet{Rodriguez2013}: (1) a circumstellar disk around DF~Cyg itself, and (2) a circumstellar disk around a companion star, and this companion/disk system periodically occulting DF Cyg. However, neither of these is geometrically convincing. First, to not occult DF~Cyg during half of the cycle but then occult $\sim$90\% of DF~Cyg during the other half of the cycle, a disk around DF~Cyg would need to have a very ``tall" feature or warp on one side. Given the radius of DF~Cyg ($\approx$10.3~R$_\odot$), this warp would need to extend vertically by $\sim$10~$R_\odot$; with an inferred total disk extent of $\sim$0.2~AU, or $\sim$50~R$_\odot$, this would be a remarkably large perturbation indeed. On the other hand, to periodically occult DF~Cyg with a period equal to the orbital period of the companion star, and assuming a Keplerian disk, the disk warp would need to be located at the same distance as the companion star. However, it would then presumably not be possible for the disk to remain stable if it completely fills the binary orbit. 

The second possibility, in which the binary companion is encircled by a dusty disk, and it is the binary companion/disk system that periodically occults DF~Cyg at the binary orbital period, is also geometrically impossible. Specifically, based on the fraction of the total light curve duration that the dimming event spans, the binary companion and its disk would need to spend roughly half of the orbital period in front of DF~Cyg, leaving half of the orbital period for the binary to traverse all other orbital phases. Even imagining that the companion is on a highly eccentric orbit and that the orbit is moreover fortuitously aligned with the observer such that the occultations correspond to apastron, it is not geometrically possible to explain the occultation durations relative to the entire orbit in such a scenario.

Thus, we are led to conclude that the opaque screen periodically occulting DF~Cyg is a circumbinary dusty disk around an entire binary system. This would represent an interpretation similar to that described in \citet{Gorlova2015} for the IRAS~19135+3937 interacting binary system (see their Figure 18), although much simpler in this case as there is no evidence in DF~Cyg for the additional complexities of accretion, jets, or any reflection effects.

By adopting the interpretation that an opaque disk periodically occults DF~Cyg, we therefore infer that the 2MASS $JHK_S$ measurements, which were obtained during the long-period minimum (Section~\ref{sec:sed}), are a combination of infrared flux from the disk plus the unobscured part of the photosphere of DF~Cyg\footnote{We ignore any contribution from the binary companion, which is presumably very faint in the visible in comparison to DF~Cyg. This assumption is corroborated by the fact that the radial velocity monitoring of DF~Cyg that identified the orbital motion of DF~Cyg did not observe the spectrum of the companion \citep{Gorlova2013}.}. If we assume that the stellar SED model in Figure~\ref{fig:sed} represents the expected flux from the full photosphere, we can use the 2MASS measurements to isolate the $JHK_S$ flux of the disk alone. To do this, we estimated the ``undimmed" 2MASS $JHK_S$ fluxes by interpolating from a straight line in Figure~\ref{fig:sed} between the 0.8 and 3.5~$\mu$m measurements. The resulting ratio of the dimmed to the undimmed flux at $JHK_S$ is $\sim$0.19, $\sim$0.22, and $\sim$0.33, respectively. 

These ratios indicate greater dimming at shorter wavelengths, which could be interpreted as interstellar-like extinction if the occulting disk is semi-transparent. Therefore, we compared these flux ratios, as a function of wavelength, to an extinction model based on the \citet{Cardelli1989} interstellar extinction law. In this comparison, we normalized the extinction law to the $\sim$0.10 flux ratio we measured from DF Cyg's {\it Kepler} light curve, adopting the midpoint of 0.66~$\mu$m for the {\it Kepler} bandpass. We found that the $JHK_S$ flux ratios are dimmer than that expected from the interstellar extinction law. Alternatively, if the occulting disk is opaque and black (i.e., non-emitting), we would expect the dimming to be gray. However, the flux ratios measured above are not gray. Rather, the trend from the {\it Kepler} and $JHK_S$ flux ratios falls in between the two extreme cases of interstellar-like extinction and an opaque non-emitting disk,  suggesting that the occulting screen is opaque but also glowing in the near-IR.

\begin{figure}%[!ht]
\centering
\includegraphics[width=0.65\textwidth,height=0.55\textheight,keepaspectratio]{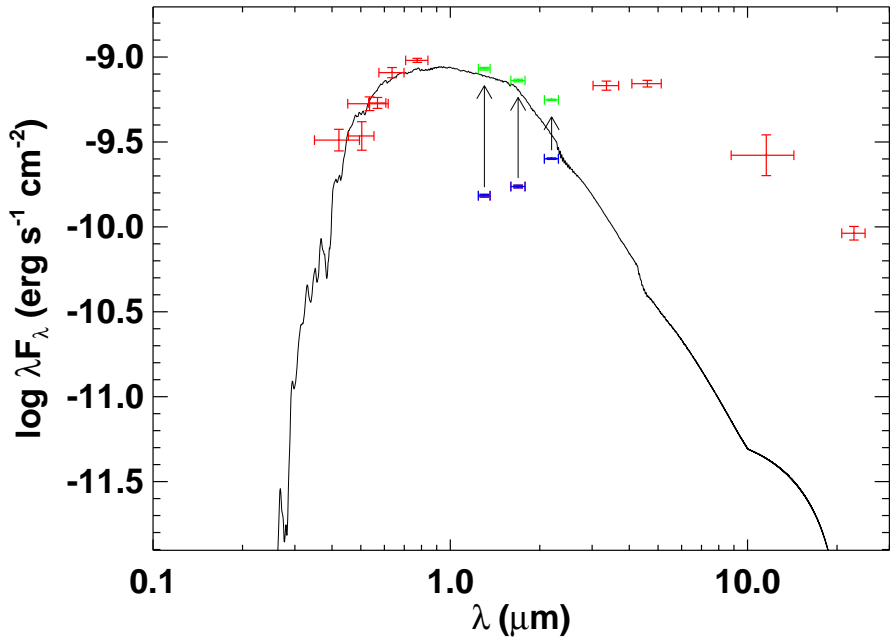}
\caption{Same as Figure~\ref{fig:sed}, but now including the $JHK_S$ fluxes corrected upward (arrows pointing to green symbols) for the obscuration of the luminous, occulting disk (see the text).}\label{fig:sedcorr}
\end{figure}

To test for a glowing disk, we attempted to estimate how bright the 2MASS data points would have been had they been measured at the long-period maximum state. If the observed 2MASS measurements come from 10\% of the DF~Cyg photosphere and 100\% of the disk, the relation between the fluxes is described by 
\begin{equation}
    F_{2MASS} = 0.1 F_{DF Cyg}    +    F_{Disk},
\end{equation}
\noindent and the total flux expected at the long-period maximum is given by
\begin{equation}
    F_{Total} = F_{DF Cyg}    +    F_{Disk}.
\end{equation}
\noindent Here, $F_{2MASS}$ is the 2MASS flux measurement, $F_{DF Cyg}$ is the flux of DF~Cyg, which we estimated from our SED photosphere model, and $F_{Disk}$ is the flux of the disk. By solving equations 1 and 2 for each 2MASS $JHK_S$ bandpass individually, we find total fluxes of $10^{-9.0}$, $10^{-9.09}$, and $10^{-9.2}$ ${\rm ~erg~s}^{-1} {\rm ~cm}^{-2}$, respectively. These values increase the originally measured 2MASS points, as shown in Figure~\ref{fig:sedcorr}, now revealing a double-peaked SED. Such a double-peaked SED is similar to that of other RV Tau stars with disks \citep[see, e.g.,][]{Deruyter2005RVTau}, lending additional credibility to the occulting disk scenario. From the shape of the SED, we estimate that the disk peaks at $\sim3.5-4\mu m$. Using Wien's displacement law we can then estimate the temperature (of the inner edge) of the disk to be $\sim$770~K.

Studies such by \citet[]{Deruyter2005RVTau,DeRuyter2006PostAGBs} have previously suggested that dusty disks, possibly circumbinary disks, around RV Tau and post-AGB stars can be stable over long timescales.
A few recent studies provide more compelling evidence for dusty circumbinary disks around post-AGB and RV Tau stars \cite[]{Gielen2009A,Hillen2014,Hillen2015ACHer}. \cite{Gielen2009A} used SED modeling of four RV Tau stars and inferred dusty disks around them. They moreover found that the disks were likely to be puffed up, i.e., having large scale heights. They also determined that these disks possess inner holes, where the inner disk radii are typically a few astronomical units, similar to the semimajor axis that we derive above for the DF Cyg binary orbit. 

\cite{Hillen2014} used interferometric measurements and SED fitting to similarly infer a puffed-up circumbinary disk with an inner hole around the binary post-AGB star 89~Her. 
The inferred disk scale height in that system suggests that a partial blocking of the central star should occur for $\sim$20-30\% of viewing angles. This fraction is also consistent with occurrence rates of RVb stars. For example, GCVS currently has 159 RV Tau stars on record, 45 classified as RVa's and 19 classified as RVb's (the rest are not listed with a classification), giving an RVb occurrence of 30\% among the classified systems. 

Note that this interpretation requires that some RVa-type stars also have disks but that these be viewed from more pole-on orientations such that the central RV Tau star is not obscured. \citet{Vanwinckel1999} was the first to suggest viewing angle and variable circumstellar extinction determined the RV~Tau's photometric class. Indeed there are known examples of such systems. For example, AC~Her, an RVa pulsator (no long-period dimming events) is in a confirmed binary system \citep{VanWinckel1998ACHER} surrounded by a stable circumbinary disk. \citet{Hillen2015ACHer} recently found evidence that AC~Her's circumbinary disk is puffed up with a large inner disk radius. 

%\newpage
%----------------------------------------------------------------------------------------
% 	Section:  SUMMARY & CONCLUSIONS
%----------------------------------------------------------------------------------------
\section{Summary and Conclusions} \label{sec:summary}

We have studied the {\it Kepler} light curve of the RV Tau variable DF Cyg with unprecedented photometric and temporal resolution. We determined that the characteristic RV Tau variable short-term pulsation has a period of \shortper~days and the RVb longer-term variation has a period of \longper~days, which are consistent with period values in the literature. We found that the overall mean flux decreases by $\sim$90\% from the long-period maxima to the long-period minimum and that the short-period peak-to-peak amplitudes, during long-period maxima, also decrease by $\sim$90\% at long-period minimum. 

During the maxima of the long-period behavior, we identified two temporary interruptions of the deep and shallow minima occurring on the order of $\sim$800~days. Studying the residuals of the arrival times for the deep minima, we discovered a similar trend on the order of $\sim$800~days. A previously measured radial velocity period of $\sim$775~days suggests an orbital period consistent with the $\sim$800~day features present in the light curve. 

Hence, there is compelling and corroborative evidence to associate the long-period variation and other $\sim$800~day features with a binary orbital period. 
It has also been suggested that the long-period behavior in RVb stars is due to the geometry of circumstellar disks where the stars are periodically occulted by their disks. Using DF Cyg's stellar parameters, based on the recent {\it Gaia} parallax measurement, and our analysis of the {\it Kepler} light curve, we considered such a scenario for DF Cyg. 

We argue that if the periodic long-term behavior is due to orbital modulation by an opaque body that blocks $\sim90\%$ DF Cyg + companion's light, then the only feasible configuration is a puffed-up circumbinary dusty disk surrounding the DF Cyg system. 
Such a scenario is consistent with DF Cyg's IR excess, which suggests the presence of a dusty disk. This interpretation is bolstered by our analysis of the 2MASS $JHK_S$ observations (taken during long-period minimum), which indicates the presence of a double-peaked SED in the near-IR similar to that of other RV Tau stars. This, in turn, suggests that the disk occultation scenario advanced here may apply generally to RVb-type variables.

\acknowledgements
L.D.V., K.G.S., and R.M.\ acknowledge partial support from NSF PAARE grant AST-1358862. 
L.D.V.\ acknowledges the support of the NASA MUREP Harriett G. Jenkins Fellowship. 
G.E.S.\ acknowledges the support of the Vanderbilt Office of the Provost through the Vanderbilt Initiative in Data-intensive Astrophysics (VIDA) fellowship.
This paper includes data collected by the Kepler mission. Funding for the Kepler mission is provided by the NASA Science Mission directorate. The data was obtained from the Mikulski Archive for Space Telescopes (MAST). STScI is operated by the Association of Universities for Research in Astronomy, Inc., under NASA contract NAS5-26555. Support for MAST for non-HST data is provided by the NASA Office of Space Science via grant NNX09AF08G and by other grants and contracts. This work has also made use of data from the European Space Agency (ESA) mission {\it Gaia} (\url{http://www.cosmos.esa.int/gaia}), processed by the {\it Gaia} Data Processing and Analysis Consortium (DPAC, \url{http://www.cosmos.esa.int/web/gaia/dpac/consortium}). Funding for the DPAC has been provided by national institutions, in particular
the institutions participating in the {\it Gaia} Multilateral Agreement. This publication makes use of data products from the Two Micron All Sky Survey, which is a joint project of the University of Massachusetts and the Infrared Processing and Analysis Center/California Institute of Technology, funded by the National Aeronautics and Space Administration and the National Science Foundation. This publication makes use of data products from the {\it Wide-field Infrared Survey Explorer}, which is a joint project of the University of California, Los Angeles, and the Jet Propulsion Laboratory/California Institute of Technology, funded by the National Aeronautics and Space Administration.
 
\facilities{{\it Gaia}, {\it GALEX}, {\it Kepler}, {\it WISE}}

%----------------------------------------------------------------------------------------
% 	Section:  REFERENCES
%----------------------------------------------------------------------------------------
\bibliography{main.bbl}

\begin{thebibliography}{}
\expandafter\ifx\csname natexlab\endcsname\relax\def\natexlab#1{#1}\fi

\bibitem[{{B{\'o}di} {et~al.}(2016){B{\'o}di}, {Szatm{\'a}ry}, \&
  {Kiss}}]{Bodi2016}
{B{\'o}di}, A., {Szatm{\'a}ry}, K., \& {Kiss}, L.~L. 2016, \aap, 596, A24

\bibitem[{{Brown} {et~al.}(2011){Brown}, {Latham}, {Everett}, \&
  {Esquerdo}}]{Brown2011}
{Brown}, T.~M., {Latham}, D.~W., {Everett}, M.~E., \& {Esquerdo}, G.~A. 2011,
  \aj, 142, 112

\bibitem[{{Cardelli} {et~al.}(1989){Cardelli}, {Clayton}, \&
  {Mathis}}]{Cardelli1989}
{Cardelli}, J.~A., {Clayton}, G.~C., \& {Mathis}, J.~S. 1989, \apj, 345, 245

\bibitem[{{De Marco}(2009)}]{Demarco2009}
{De Marco}, O. 2009, \pasp, 121, 316

\bibitem[{{de Ruyter} {et~al.}(2005){de Ruyter}, {van Winckel}, {Dominik},
  {Waters}, \& {Dejonghe}}]{Deruyter2005RVTau}
{de Ruyter}, S., {van Winckel}, H., {Dominik}, C., {Waters}, L.~B.~F.~M., \&
  {Dejonghe}, H. 2005, \aap, 435, 161

\bibitem[{{de Ruyter} {et~al.}(2006){de Ruyter}, {van Winckel}, {Maas}, {Lloyd
  Evans}, {Waters}, \& {Dejonghe}}]{DeRuyter2006PostAGBs}
{de Ruyter}, S., {van Winckel}, H., {Maas}, T., {et~al.} 2006, \aap, 448, 641

\bibitem[{{Evans}(1985)}]{LloydEvans1985}
{Evans}, T.~L. 1985, \mnras, 217, 493

\bibitem[{{Gaia Collaboration} {et~al.}(2016{\natexlab{a}}){Gaia
  Collaboration}, {Brown}, {Vallenari}, {Prusti}, {de Bruijne}, {Mignard},
  {Drimmel}, {Babusiaux}, {Bailer-Jones}, {Bastian}, \& et~al.}]{Gaia2016a}
{Gaia Collaboration}, {Brown}, A.~G.~A., {Vallenari}, A., {et~al.}
  2016{\natexlab{a}}, \aap, 595, A2

\bibitem[{{Gaia Collaboration} {et~al.}(2016{\natexlab{b}}){Gaia
  Collaboration}, {Prusti}, {de Bruijne}, {Brown}, {Vallenari}, {Babusiaux},
  {Bailer-Jones}, {Bastian}, {Biermann}, {Evans}, \& et~al.}]{Gaia2016b}
{Gaia Collaboration}, {Prusti}, T., {de Bruijne}, J.~H.~J., {et~al.}
  2016{\natexlab{b}}, \aap, 595, A1

\bibitem[{{Gerasimovi{\v c}}(1929)}]{Gerasimovic1929}
{Gerasimovi{\v c}}, B.~P. 1929, Harvard College Observatory Circular, 341, 1

\bibitem[{{Gezer} {et~al.}(2015){Gezer}, {Van Winckel}, {Bozkurt}, {De Smedt},
  {Kamath}, {Hillen}, \& {Manick}}]{Gezer2015}
{Gezer}, I., {Van Winckel}, H., {Bozkurt}, Z., {et~al.} 2015, \mnras, 453, 133

\bibitem[{{Gielen} {et~al.}(2009{\natexlab{a}}){Gielen}, {van Winckel},
  {Matsuura}, {Min}, {Deroo}, {Waters}, \& {Dominik}}]{Gielen2009}
{Gielen}, C., {van Winckel}, H., {Matsuura}, M., {et~al.} 2009{\natexlab{a}},
  \aap, 503, 843

\bibitem[{{Gielen} {et~al.}(2009{\natexlab{b}}){Gielen}, {van Winckel},
  {Reyniers}, {Zijlstra}, {Lloyd Evans}, {Gordon}, {Kemper}, {Indebetouw},
  {Marengo}, {Matsuura}, {Meixner}, {Sloan}, {Tielens}, \&
  {Woods}}]{Gielen2009A}
{Gielen}, C., {van Winckel}, H., {Reyniers}, M., {et~al.} 2009{\natexlab{b}},
  \aap, 508, 1391

\bibitem[{{Giridhar} {et~al.}(2005){Giridhar}, {Lambert}, {Reddy}, {Gonzalez},
  \& {Yong}}]{Giridhar2005}
{Giridhar}, S., {Lambert}, D.~L., {Reddy}, B.~E., {Gonzalez}, G., \& {Yong}, D.
  2005, \apj, 627, 432

\bibitem[{{Gorlova} {et~al.}(2013){Gorlova}, {Van Winckel}, {Vos},
  {{\O}stensen}, {Jorissen}, {Van Eck}, \& {Ikonnikova}}]{Gorlova2013}
{Gorlova}, N., {Van Winckel}, H., {Vos}, J., {et~al.} 2013, in EAS Publications
  Series, Vol.~64, Setting a New Standard in the Analysis of Binary Stars, ed.
  K.~{Pavlovski}, A.~{Tkachenko}, \& G.~{Torres}, 163

\bibitem[{{Gorlova} {et~al.}(2015){Gorlova}, {Van Winckel}, {Ikonnikova},
  {Burlak}, {Komissarova}, {Jorissen}, {Gielen}, {Debosscher}, \&
  {Degroote}}]{Gorlova2015}
{Gorlova}, N., {Van Winckel}, H., {Ikonnikova}, N.~P., {et~al.} 2015, \mnras,
  451, 2462

\bibitem[{{Green} {et~al.}(2015){Green}, {Schlafly}, {Finkbeiner}, {Rix},
  {Martin}, {Burgett}, {Draper}, {Flewelling}, {Hodapp}, {Kaiser}, {Kudritzki},
  {Magnier}, {Metcalfe}, {Price}, {Tonry}, \& {Wainscoat}}]{Green:2015}
{Green}, G.~M., {Schlafly}, E.~F., {Finkbeiner}, D.~P., {et~al.} 2015, \apj,
  810, 25

\bibitem[{{Hartig} {et~al.}(2014){Hartig}, {Cash}, {Hinkle}, {Lebzelter},
  {Mighell}, \& {Walter}}]{Hartig2014}
{Hartig}, E., {Cash}, J., {Hinkle}, K.~H., {et~al.} 2014, \aj, 148, 123

\bibitem[{{Harwood}(1927)}]{Harwood1927}
{Harwood}, M. 1927, Harvard College Observatory Bulletin, 847, 5

\bibitem[{{Harwood} \& {Shapley}(1937)}]{Harwood1937}
{Harwood}, M., \& {Shapley}, H. 1937, Annals of Harvard College Observatory,
  105, 521

\bibitem[{{Hillen} {et~al.}(2015){Hillen}, {de Vries}, {Menu}, {Van Winckel},
  {Min}, \& {Mulders}}]{Hillen2015ACHer}
{Hillen}, M., {de Vries}, B.~L., {Menu}, J., {et~al.} 2015, \aap, 578, A40

\bibitem[{{Hillen} {et~al.}(2014){Hillen}, {Menu}, {Van Winckel}, {Min},
  {Gielen}, {Wevers}, {Mulders}, {Regibo}, \& {Verhoelst}}]{Hillen2014}
{Hillen}, M., {Menu}, J., {Van Winckel}, H., {et~al.} 2014, \aap, 568, A12

\bibitem[{{Joy}(1952)}]{Joy1952}
{Joy}, A.~H. 1952, \apj, 115, 25

\bibitem[{{Jura}(1986)}]{Jura1986}
{Jura}, M. 1986, \apj, 309, 732

\bibitem[{{Maas} {et~al.}(2002){Maas}, {Van Winckel}, \& {Waelkens}}]{Maas2002}
{Maas}, T., {Van Winckel}, H., \& {Waelkens}, C. 2002, \aap, 386, 504

\bibitem[{{Manick} {et~al.}(2017){Manick}, {Van Winckel}, {Kamath}, {Hillen},
  \& {Escorza}}]{Manick2016}
{Manick}, R., {Van Winckel}, H., {Kamath}, D., {Hillen}, M., \& {Escorza}, A.
  2017, \aap, 597, A129

\bibitem[{{Percy}(1993)}]{Percy1993}
{Percy}, J.~R. 1993, in Astronomical Society of the Pacific Conference Series,
  Vol.~45, Luminous High-Latitude Stars, ed. D.~D. {Sasselov}, 295

\bibitem[{{Percy} {et~al.}(1991){Percy}, {Sasselov}, {Alfred}, \&
  {Scott}}]{Percy1991}
{Percy}, J.~R., {Sasselov}, D.~D., {Alfred}, A., \& {Scott}, G. 1991, \apj,
  375, 691

\bibitem[{{Pollard} \& {Cottrell}(1995)}]{Pollard1995}
{Pollard}, K.~H., \& {Cottrell}, P.~L. 1995, in Astronomical Society of the
  Pacific Conference Series, Vol.~83, IAU Colloq. 155: Astrophysical
  Applications of Stellar Pulsation, ed. R.~S. {Stobie} \& P.~A. {Whitelock},
  409

\bibitem[{{Pollard} {et~al.}(1996){Pollard}, {Cottrell}, {Kilmartin}, \&
  {Gilmore}}]{Pollard1996}
{Pollard}, K.~R., {Cottrell}, P.~L., {Kilmartin}, P.~M., \& {Gilmore}, A.~C.
  1996, \mnras, 279, 949

\bibitem[{{Rodriguez} {et~al.}(2013){Rodriguez}, {Pepper}, {Stassun}, {Siverd},
  {Cargile}, {Beatty}, \& {Gaudi}}]{Rodriguez2013}
{Rodriguez}, J.~E., {Pepper}, J., {Stassun}, K.~G., {et~al.} 2013, \aj, 146,
  112

\bibitem[{{Samus} {et~al.}(2009){Samus}, {Durlevich}, \& {et al.}}]{Samus2009}
{Samus}, N.~N., {Durlevich}, O.~V., \& {et al.} 2009, yCat, 1, 2025

\bibitem[{{Skrutskie} {et~al.}(2006){Skrutskie}, {Cutri}, {Stiening},
  {Weinberg}, {Schneider}, {Carpenter}, {Beichman}, {Capps}, {Chester},
  {Elias}, {Huchra}, {Liebert}, {Lonsdale}, {Monet}, {Price}, {Seitzer},
  {Jarrett}, {Kirkpatrick}, {Gizis}, {Howard}, {Evans}, {Fowler}, {Fullmer},
  {Hurt}, {Light}, {Kopan}, {Marsh}, {McCallon}, {Tam}, {Van Dyk}, \&
  {Wheelock}}]{Skrutskie2006}
{Skrutskie}, M.~F., {Cutri}, R.~M., {Stiening}, R., {et~al.} 2006, \aj, 131,
  1163

\bibitem[{{Thompson} {et~al.}(2012){Thompson}, {Everett}, {Mullally},
  {Barclay}, {Howell}, {Still}, {Rowe}, {Christiansen}, {Kurtz}, {Hambleton},
  {Twicken}, {Ibrahim}, \& {Clarke}}]{Thompson2012}
{Thompson}, S.~E., {Everett}, M., {Mullally}, F., {et~al.} 2012, \apj, 753, 86

\bibitem[{{Tsesevich}(1975)}]{Tsesevich1975}
{Tsesevich}, V.~P. 1975, {Pulsating Stars}, ed. B.~V. {Kukarkin} (New York:
  Wiley), 112

\bibitem[{{Van Winckel} {et~al.}(1999){Van Winckel}, {Waelkens}, {Fernie}, \&
  {Waters}}]{Vanwinckel1999}
{Van Winckel}, H., {Waelkens}, C., {Fernie}, J.~D., \& {Waters}, L.~B.~F.~M.
  1999, \aap, 343, 202

\bibitem[{{Van Winckel} {et~al.}(1998){Van Winckel}, {Waelkens}, {Waters},
  {Molster}, {Udry}, \& {Bakker}}]{VanWinckel1998ACHER}
{Van Winckel}, H., {Waelkens}, C., {Waters}, L.~B.~F.~M., {et~al.} 1998, \aap,
  336, L17

\bibitem[{{Wallerstein}(2002)}]{Wallerstein2002}
{Wallerstein}, G. 2002, \pasp, 114, 689

\bibitem[{{Weidemann}(1990)}]{Weidemann1990}
{Weidemann}, V. 1990, \araa, 28, 103

\end{thebibliography}

\end{document}